\documentclass[10pt]{revtex4}
\usepackage{graphicx}
\usepackage{dcolumn}
\usepackage{bm}

\begin{document}

\preprint{}
\title{$\Delta$, $K^*$ and $\rho$ Resonance Production And Their\\
Probing of Freeze-out Dynamics at RHIC}
\author{Haibin Zhang\\for the STAR Collaboration}
\address{Brookhaven National Laboratory and Yale University}
\date{March 10th, 2004}

\begin{abstract}
We report the measurements of the transverse momentum spectra and
invariant mass distributions of $\Delta$(1232)$\rightarrow \pi p$,
$K^*$(892)$^{0,\pm} \rightarrow \pi K$ and $\rho$(770)$^{0}
\rightarrow \pi^{+} \pi^{-}$ in Au+Au and p+p collisions at
$\sqrt{s_{NN}}$=200 GeV using the STAR TPC at RHIC. These
resonances provide sensitive probes to examine the evolution
dynamics in the hadronic medium through their decay and
regeneration processes. The particle ratios of $K^*/K$,
$K^*/\phi$, $\Delta$/p and $\rho/\pi$, the $K^*$, $\rho$ and
$\Delta$ apparent masses and the dependence of these quantities on
centrality provide evidence of dynamical interaction and
re-scattering between hadrons close to freeze-out. The dependence
of resonance yields on the strength of their hadronic cross
sections and the lifetime between the freeze-outs will also be
discussed. In the intermediate $p_{T}$ region ($2<p_T<4$ GeV/c),
the nuclear modification factors ($R_{CP}$ and $R_{AA}$) of $K^*$
are similar to those of $K_{S}$ and smaller than those of baryons
($\Lambda$). In the intermediate $p_{T}$ region, no strong
dependence on the particle masses is seen in the data.

\end{abstract}
\maketitle

\section{Physics}
Resonances are strongly decaying particles which have lifetimes of
about a few fm/c. Compared to other stable particles (such as
$\Lambda$, $D^0$, etc.), resonances have unique characteristics to
probe various properties of the hot dense matter produced in
relativistic heavy ion collisions. First, in a hot dense matter,
resonances are in close encounter with other strongly interacting
hadrons. These resonances in-medium effects related to the high
density and/or high temperature of the medium can modify various
resonance properties, such as masses, widths and even the mass
line shapes~\cite{medium}. Resonances extremely short lifetimes
will enable us to directly measure these in-medium effects.
Second, resonances can decay between the chemical and kinetic
freeze-outs of the fire ball and their decayed daughters can
undergo a period of interaction with the hadrons in the medium.
This resonance decayed daughters re-scattering effect can destroy
part of the primordial resonance yields~\cite{urqmd1}. Hadrons in
the medium can interact with each other to re-generate resonance
signals~\cite{urqmd1}. Thus measuring resonance yields and their
ratios to corresponding stable particles in relativistic heavy ion
collisions compared to elementary p+p collisions will enable us to
probe the dynamics between the chemical and kinetic freeze-outs.
Third, some resonances, such as $K^*$, $\phi$, etc., are heavy
mesons with their masses close to baryons. Thus we can study
various physics topics through these heavy meson resonances, such
as to identify whether the nuclear modification factor in the
intermediate $p_T$ region depends on mass or particle species
(i.e. meson/baryon).

\section{Measurements}
During the second RHIC run (2001-2002), RHIC (the Relativistic
Heavy Ion Collider) performed Au+Au and p+p collisions at
$\sqrt{s_{NN}}$=200 GeV. In the STAR (the Solenoidal Tracker at
RHIC) detector, the Au+Au collision minimum bias trigger was
defined by requiring coincidences between two ZDCs (Zero Degree
Calorimeter) which measured the spectator neutrons. The Au+Au
collision central trigger was defined using the scintillator CTB
(Central Trigger Barrel) and both the ZDCs. The p+p collision
minimum bias trigger was defined using coincidences between two
BBCs (Beam-Beam Counter) that measured the charged particles
multiplicity near beam rapidity. In this analysis, about 2M Au+Au
minimum bias triggered, 2M Au+Au central triggered and 6M p+p
minimum bias triggered collisions events are used, which were
taken mainly using the TPC (Time Projection Chamber) in the STAR detector.\\ \\
Through the ionization energy loss ($dE/dx$) in the TPC, charged
pions and kaons are identified with momentum up to about 0.75
GeV/c, protons and anti-protons are identified with momentum up to
about 1.1 GeV/c. The $\rho^0$(770) resonance invariant mass
spectra are reconstructed using the like-sign technique~\cite{rho}
via the decay channel $\rho^0\rightarrow\pi^+\pi^-$. The
$K^*$(892) and $\Delta^{++}$(1232) resonances invariant mass
spectra are reconstructed using the event-mixing
technique~\cite{kstar130} via the decay channels
$K^{*0}\rightarrow K^+\pi^-$, $K^{*\pm}\rightarrow K_S^0\pi^{\pm}$
and $\Delta^{++}\rightarrow p\pi^+$, respectively. The $K_S^0$
signals used in the $K^{*\pm}$ reconstruction are measured from a
decay topology method via $K_S^0\rightarrow\pi^+\pi^-$.

\section{Results}
From the reconstructed $\rho^0$(770), $K^{*}$(892) and
$\Delta^{++}$(1232) resonances invariant mass spectra, a
relativistic Breit-Wigner function multiplied by a phase space
factor is used to fit with the resonance signals and the resonance
masses are extracted from the fit. Figure 1 shows the measured
$\rho^0$ and $K^{*0}$ masses in p+p and Au+Au collisions as a
function of $p_T$ and the $\Delta^{++}$ mass as a function of
charged hadron multiplicity in p+p and Au+Au collisions. From this
figure, we can see that a significant downward $\rho^0$ (compared
to the average $\rho^0$ mass measured in $e^+e^-$ and p+p
collisions) and $K^{*0}$ (compared to Monte Carlo studies which
have the 896.1 MeV/c$^2$ $K^{*0}$ mass input and consider the same
kinematic cuts and acceptance in real data analysis) mass shift at
low $p_T$ region has been observed and this downward mass shift is
$p_T$ dependent. These $\rho^0$ and $K^{*0}$ mass shifts at low
$p_T$ region indicate that the resonances in-medium effect may
have already modified the resonances properties and low $p_T$
resonances have less chances to escape the medium compared to
higher $p_T$ resonances so that low $p_T$ resonances properties
can be more likely to be affected by the in-medium effects. A
significant $\Delta^{++}$ (compared to the $\Delta^{++}$ mass
in~\cite{pdg}) mass shift toward low masses has also been observed
in p+p and Au+Au collisions and this mass shift becomes smaller as
the multiplicity increases while the $\rho^0$ mass shift becomes
larger as the multiplicity increases. This $\Delta^{++}$ and
$\rho^0$ mass shift multiplicity dependence can be possibly
explained by the resonances $t$-channel interaction in the medium
which might be able to push the $\rho^0$ mass down and push the
$\Delta^{++}$ mass up based on their observed masses in p+p
collisions~\cite{shuryak}. Various theoretical
calculations~\cite{medium,shuryak,rapp,kolb,ron} concerning the
resonances high temperature and/or high density related in-medium
\begin{figure}[h]
\centering
\includegraphics[height=12pc,width=14pc]{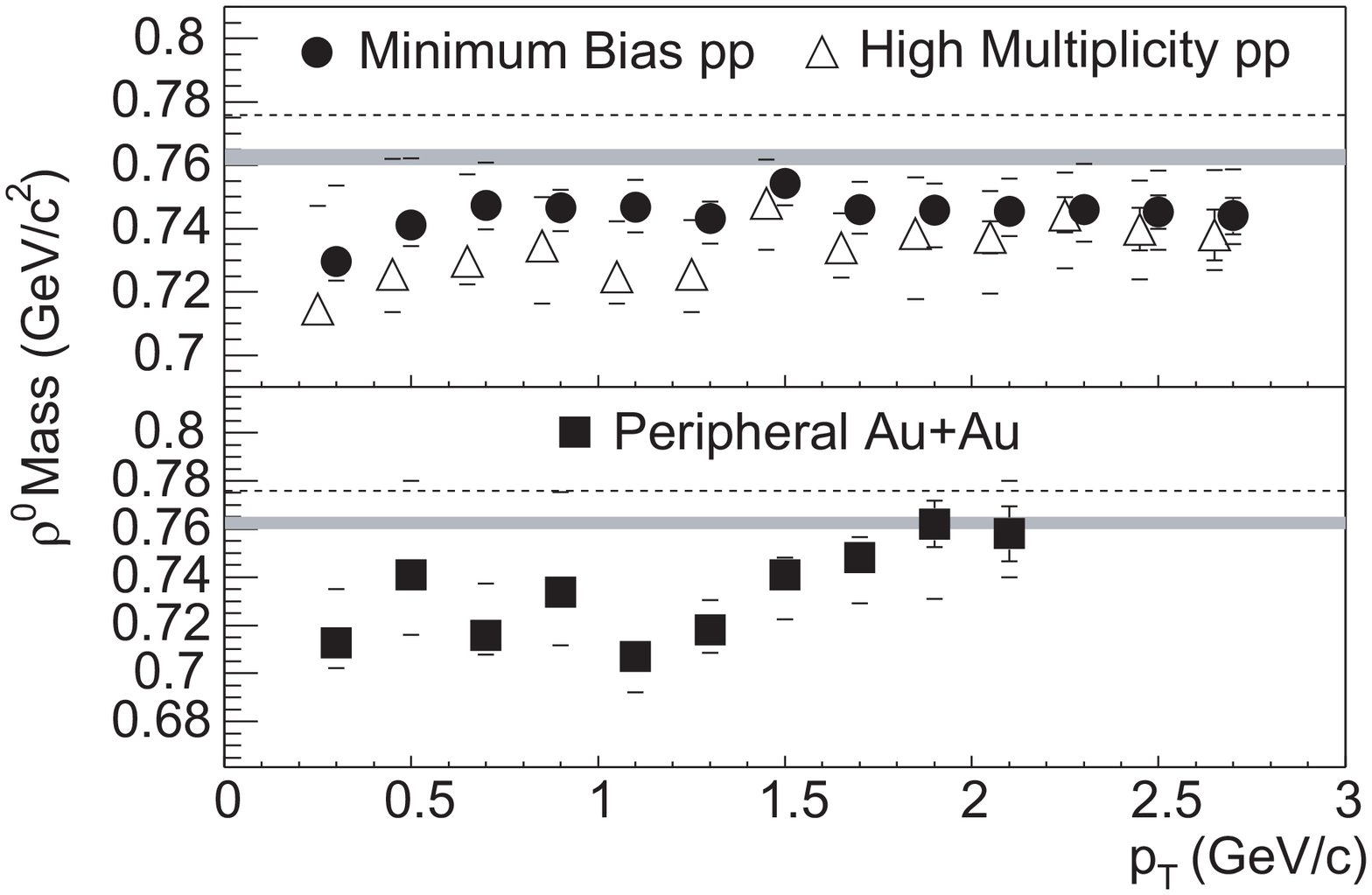}
\includegraphics[height=12pc,width=14pc]{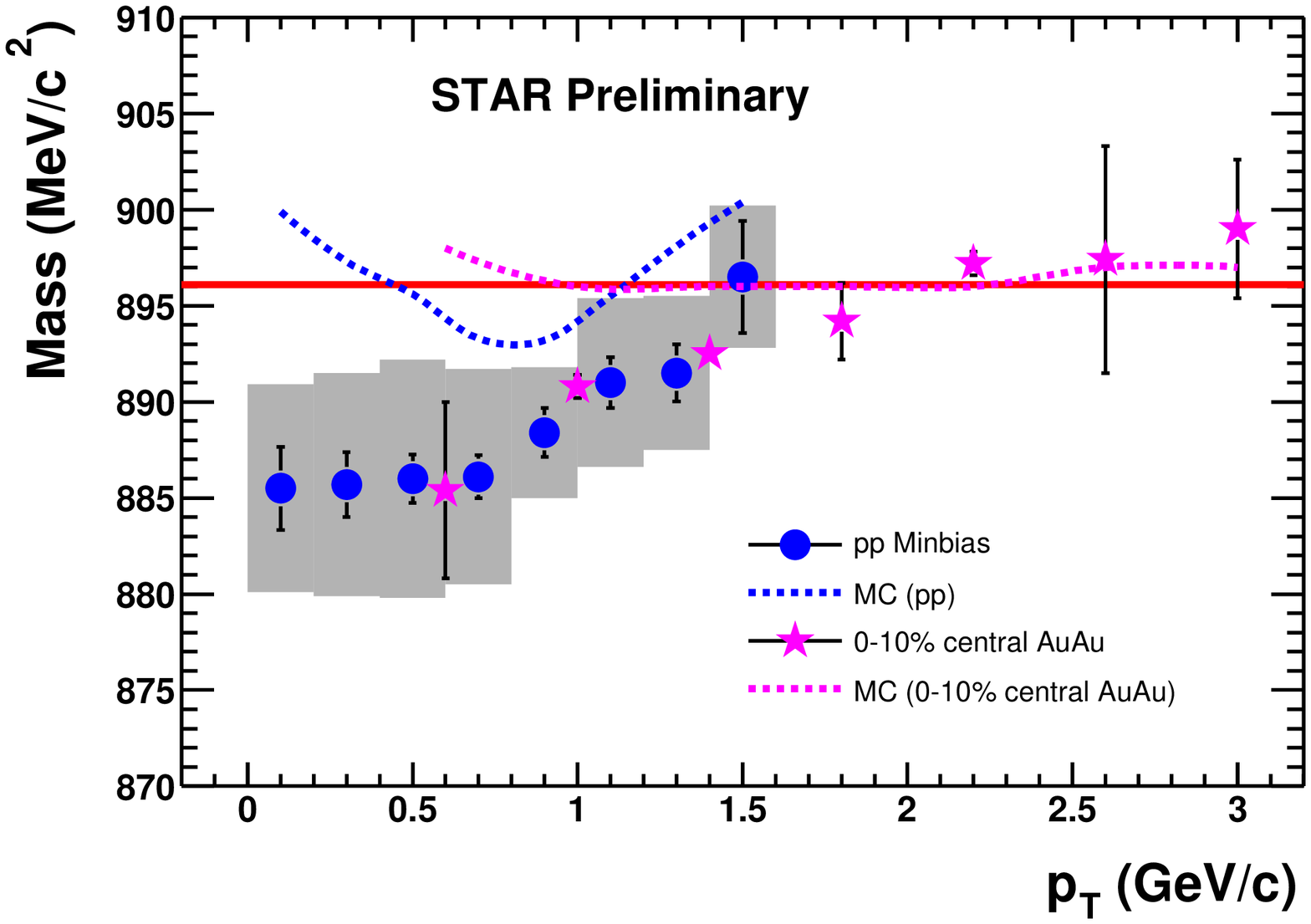}
\includegraphics[height=12pc,width=14pc]{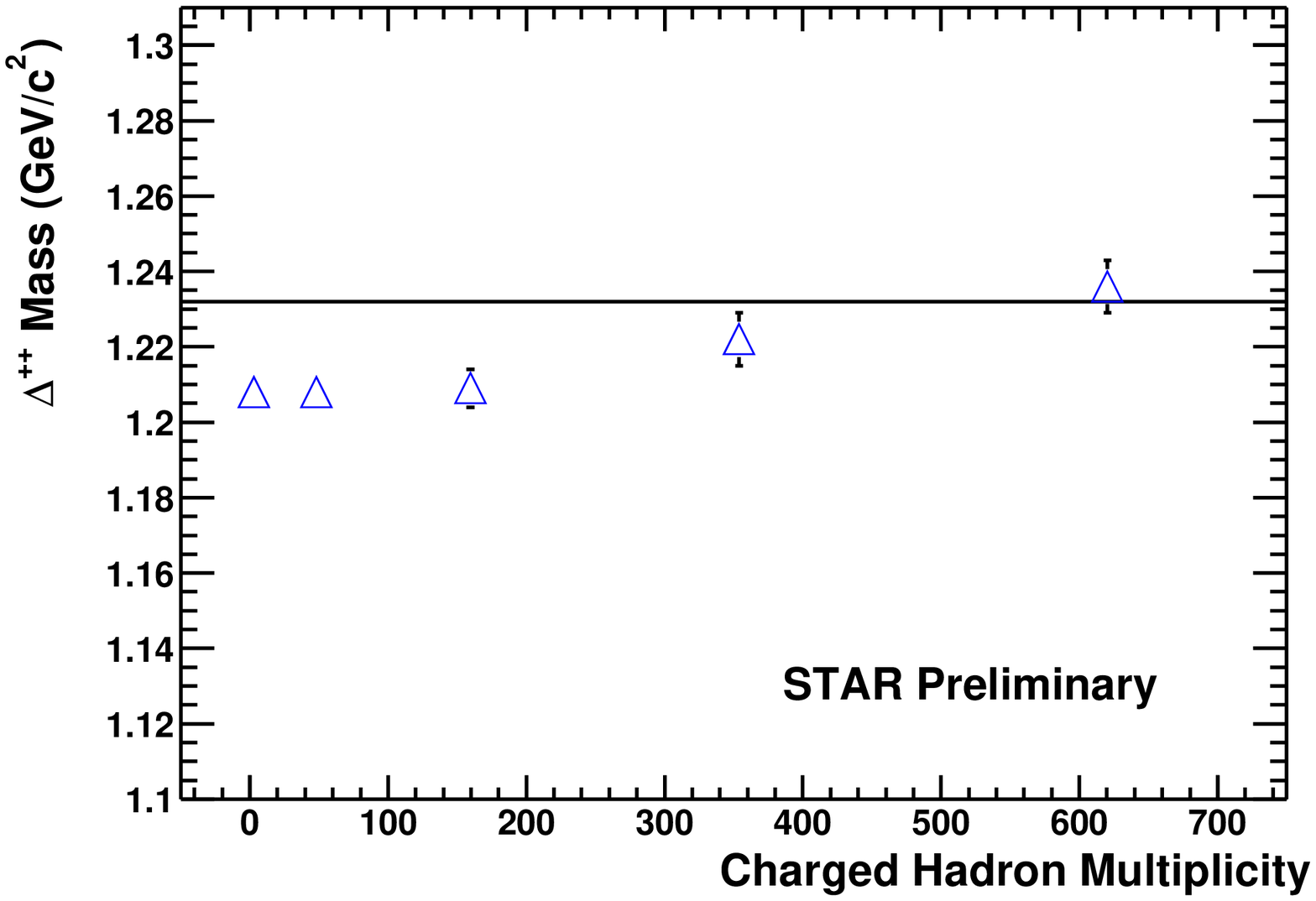}
\caption{Left: the $\rho^0$ mass as a function of $p_T$ in minimum
bias p+p, high multiplicity p+p and peripheral Au+Au collisions.
The brackets stand for systematic uncertainties. The dashed lines
represent the average of the $\rho^0$ mass measured in
$e^+e^-$~\cite{rho_e}. The shaded areas indicate the $\rho^0$ mass
measured in p+p~\cite{rho_pp}. Middle: the $K^{*0}$ mass as a
function of $p_T$ in minimum bias p+p and central Au+Au
collisions. Shadows represent systematic uncertainties in p+p. The
red line stands for the $K^{*0}$ mass 896.1 MeV/c$^2$
from~\cite{pdg}. Right: the $\Delta^{++}$ mass as a function of
charged hadron multiplicity in p+p (most left symbol) and Au+Au
collisions with different centralities (right four symbols). The
black line stand for the $\Delta^{++}$ mass 1232 MeV/c$^2$
from~\cite{pdg}.}
\end{figure}
effects also indicate similar resonances mass shifts.\\ \\
From the fit to the resonance invariant mass spectra at
mid-rapidity $|y|<0.5$ for each $p_T$ bin, the $\rho^0$, $K^{*}$
and $\Delta^{++}$ resonances raw yields can be extracted in p+p
and various centralities in Au+Au collisions. After efficiency and
acceptance correction,
\begin{figure}[h]
\centering
\includegraphics[height=18pc,width=25pc]{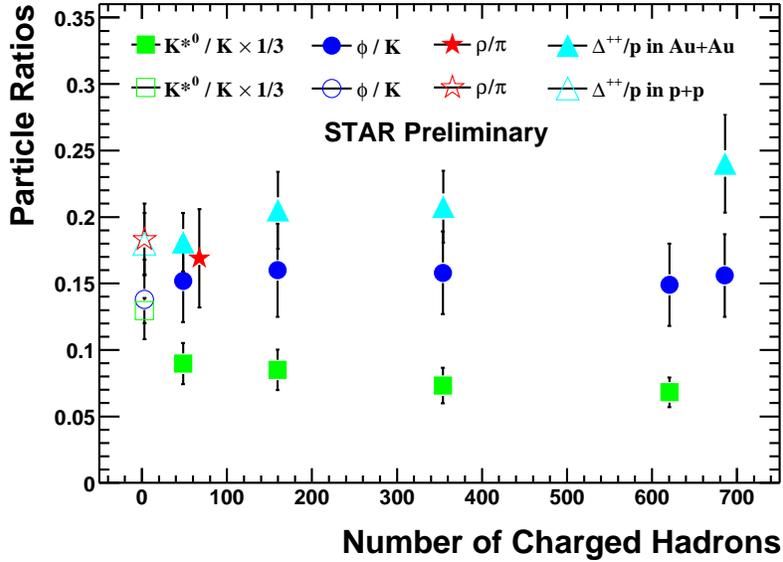}
\caption{The $K^{*0}/K$, $\rho^0/\pi$, $\Delta^{++}/p$ and
$\phi/K$ ratios as a function of number of charged hadrons in p+p
(open symbols) and various centralities in Au+Au (solid symbols)
collisions.}
\end{figure}
the above resonances invariant yield ($(2\pi)^{-1}d^2N/p_Tdp_Tdy$
or $(2\pi)^{-1}d^2N/m_Tdm_Tdy$) spectra as a function of $m_T$ (in
Au+Au collisions) or $p_T$ (in p+p collisions) can be achieved.
The $m_T$ spectra in Au+Au collisions are then fit with an
exponential function $(2\pi)^{-1}d^2N/m_Tdm_Tdy=dN/dy\times(2\pi
T(m_0+T))^{-1}\text{exp}(-(m_T-m_0)/T)$ to extract the resonances
mid-rapidity yields $dN/dy$ and the inverse slope parameters $T$.
The resonances $p_T$ spectra in p+p collisions can be well fit
with a power law function
$(2\pi)^{-1}d^2N/p_Tdp_Tdy=A\times(1+p_T/[\langle p_T\rangle
(n-3)/2])^{-n}$ indicating the hard processes at larger $p_T$
region ($p_T>2$ GeV/c) in p+p collisions.\\ \\
The stable charged hadrons ($\pi^{\pm}$, $K^{\pm}$, $p$ and
$\overline{p}$) mid-rapidity yields $dN/dy$ have also been
measured in the STAR experiment in Au+Au and p+p collisions at
$\sqrt{s_{NN}}$=200 GeV~\cite{spectra}. Thus the resonances to
their corresponding stable hadrons ratios can be calculated.
Figure 2 shows the $K^{*0}/K$, $\rho^0/\pi$, $\Delta^{++}/p$ and
$\phi/K$ ratios as a function of number of charged hadrons in p+p
and various centralities in Au+Au collisions. From this figure, we
can see that the $K^{*0}/K$ ratios in Au+Au collisions are
significantly smaller than in p+p collisions. This $K^{*0}/K$
ratio suppression in Au+Au collisions may indicate that between
the chemical freeze-out and kinetic freeze-out, more $K^{*}$
resonance signals are destroyed by the daughter particles'
re-scattering effect than the signals produced by the
re-generation effect, since the $\pi-\pi$ total interaction cross
section, which mainly determines how strong the re-scattering
effect is, is significantly larger than the $K-\pi$ total
interaction cross section~\cite{urqmd2}, which decides the
re-generation effect. In the case of the $\rho$ resonance, both
the re-scattering and re-generation effects are determined by the
$\pi-\pi$ total interaction cross section. Thus the $\rho^0/\pi$
ratios in both p+p and peripheral Au+Au collisions are comparable.
In the case of the $\Delta$ resonance, the $p-\pi$ total
interaction cross section is about 1.2 times of the $\pi-\pi$
total interaction cross~\cite{urqmd2} so that the observed
$\Delta^{++}/p$ ratios are increasing from p+p collisions to
peripheral Au+Au and to central Au+Au collisions. In the case of
the $\phi$ resonance, its lifetime is relatively larger ($\sim$40
fm/c) compared to the $\rho$, $K^*$ and $\Delta$ resonances so
that a smaller portion of the $\phi$ signals may decay before the
kinetic freeze-out. On the other hand, the $K-K$ total interaction
cross section to produce a $\phi$ meson is small~\cite{urqmd2}.
Strong re-scattering and re-generation effects are not expected in
the case of $\phi$ meson between the chemical and kinetic
freeze-outs. Thus the $\phi/K$ ratios in p+p and Au+Au
\begin{figure}[h]
\centering
\includegraphics[height=16pc,width=20pc]{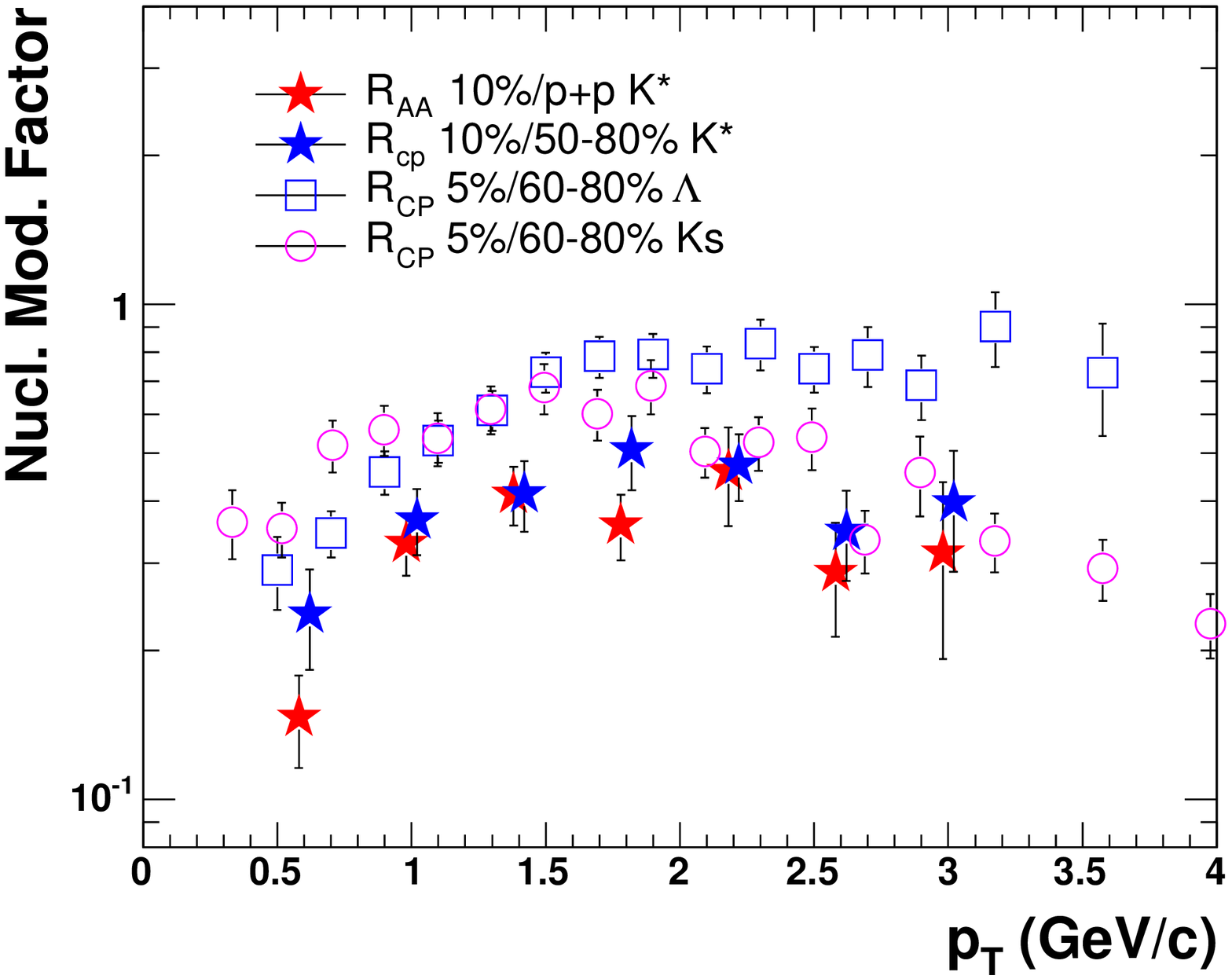}
\includegraphics[height=16pc,width=20pc]{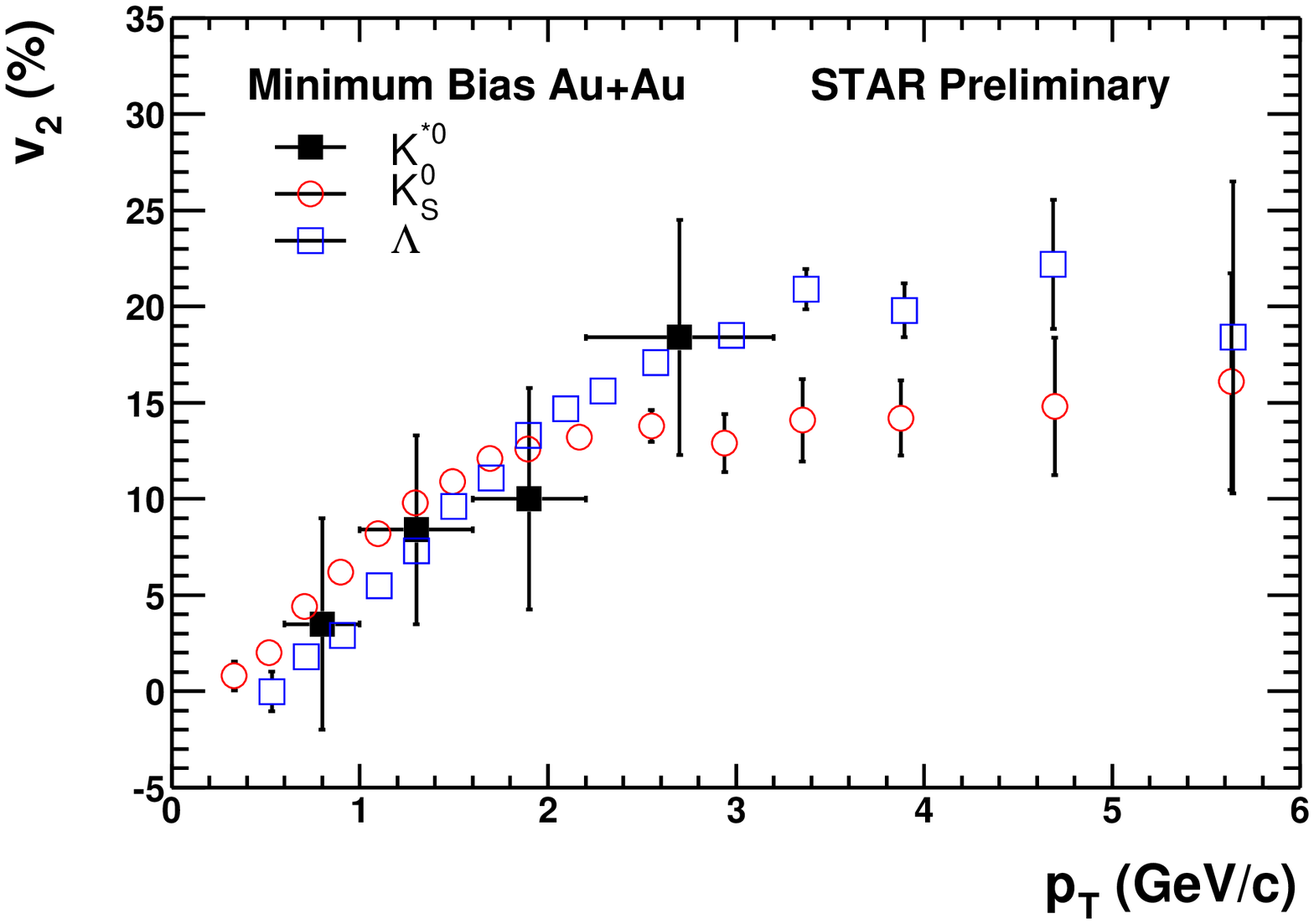}
\caption{Left: The $K^{*}$ nuclear modification factor $R_{CP}$
and $R_{AA}$ as a function of $p_T$ compared to the $\Lambda$ and
$K_S^0$ $R_{CP}$. Right: The $K^{*0}$ elliptic flow $v_2$ as a
function of $p_T$ measured in minimum bias triggered Au+Au
collisions compared to the $\Lambda$ and $K_S^0$ $v_2$. Data
points include statistical uncertainties only.}
\end{figure}
collisions are comparable to each other.\\ \\
In the STAR experiment, the $\Lambda$ and $K_S^0$ nuclear
modification factor $R_{CP}$
($=[d^2N/dp_T/d\eta/N_{coll}|_{central}]$
$/[d^2N/dp_T/d\eta/N_{coll}|_{peripheral}]$) have been
measured~\cite{rcp}. In the intermediate $p_T$ region ($1.6<p_T<4$
GeV/c), the $\Lambda$ and $K_S^0$ $R_{CP}$ are significantly
smaller than unity indicating the high $p_T$ jets lose energy
through gluon radiation while traversing through a dense matter.
It has also been observed that the nuclear modification factors
are significantly different for $\Lambda$ and $K_S^0$ with
$p_T>1.6$ GeV/c. It's important to identify whether this $R_{CP}$
difference is due to a mass effect or a particle species effect
since $\Lambda$ is a baryon and $K_S^0$ is a meson. The $K^{*}$
resonance is a meson but with its mass close to the $\Lambda$
baryon. The left panel of Figure 3 shows the $K^{*}$ nuclear
modification factor $R_{CP}$ and $R_{AA}$
($=[d^2N^{AA}/dp_Td\eta]/[T_{AA}d^2\sigma^{NN}/dp_Td\eta]$) as a
function of $p_T$ and compared to the $\Lambda$ and $K_S^0$
$R_{CP}$. From this figure, we can see that both the $K^*$
$R_{CP}$ and $R_{AA}$ at $p_T<1.6$ GeV/c are smaller than the
$\Lambda$ and $K_S^0$ $R_{CP}$ indicating the $K^*$ daughter
particles' re-scattering effect destroys low $p_T$ signals. At
$p_T>1.6$ GeV/c, the $K^*$ daughter particles' re-scattering
effect is weak since high $p_T$ $K^*$ resonances have more chances
to escape the fire ball and avoid the re-scattering effect. In
Figure 3, we can see that both the $K^*$ $R_{CP}$ and $R_{AA}$ are
closer to the $K_S^0$ $R_{CP}$ and different from the $\Lambda$
$R_{CP}$ at $p_T>1.6$ GeV/c. Thus a strong mass dependence of the
nuclear modification factor has not been observed
in this analysis.\\ \\
Identified hadron elliptic flow $v_2$ measurements have shown that
in the intermediate $p_T$ region ($2<p_T<4$ GeV/c), $v_2$ obeys
the simple scaling law $v_2(p_T)=nv_2^q(p_T/n)$, where $n$ is the
number of valence quarks of the hadron~\cite{resonance_v2}. In the
case of the $K^*$ resonance, if the $K^*$ is produced from a
hadronizing quark gluon plasma via quark recombinations, the $K^*$
$v_2$ obeys the scaling law with $n=2$; if the $K^*$ is produced
in the hadronic final state via $K-\pi$ scattering, the $K^*$
$v_2$ obeys the scaling law with $n=4$~\cite{resonance_v2}. The
final observed $K^*$ $v_2$ at the intermediate $p_T$ region should
be a combination of the above two extreme cases with certain
fractions. It is important to measure the $K^{*}$ $v_2$ and the
combination fractions to probe the resonance production mechanism
in relativistic heavy ion collisions. The right panel of Figure 3
shows the $K^{*0}$ $v_2$ as a function of $p_T$ measured in
minimum bias triggered Au+Au collisions compared to the $K_S^0$
(with $n=2$ in the scale law) and $\Lambda$ (with $n=3$ in the
scaling law) $v_2$. Due the large errors of the the measured
$K^{*0}$ $v_2$, no significant difference has been seen between
the $K^{*0}$ $v_2$ and the $v_2$ of $K_S^0$ and $\Lambda$. More
statistics is needed in the coming Au+Au run in 2004 at RHIC to
precisely measure the $K^*$ $v_2$.

\section{Conclusions}
During the second RHIC run, the $\rho^0$(770), $K^*$(892) and
$\Delta^{++}$(1232) resonances have been measured using the TPC of
the STAR detector in p+p and Au+Au collisions at
$\sqrt{s_{NN}}$=200 GeV. A downward mass shift has been observed
for the $\rho^0$ and $K^{*0}$ resonances as a function of $p_T$
and for the $\Delta^{++}$ resonance as a function of charged
hadron multiplicity in both p+p and Au+Au collisions indicating
the resonances in-medium effects might have already modified the
resonances properties in the medium. The $K^{*0}/K$, $\rho^0/\pi$
and $\Delta^{++}/p$ ratios in p+p and Au+Au collisions have been
calculated and various dynamics in the hadron medium between
chemical and kinetic freeze-outs in relativistic Au+Au collisions
have been discussed. The $K^*$ nuclear modification factor
$R_{CP}$ and $R_{AA}$ have been measured as a function of $p_T$
and in the intermediate $p_T$ region ($p_T>1.6$ GeV/c). There is
no significant mass dependence has been observed for the nuclear
modification factor by comparing the $K^*$ $R_{CP}$ and $R_{AA}$
to the $\Lambda$ and $K_S^0$ $R_{CP}$.

\end{document}